\documentclass{article}

\usepackage{PRIMEarxiv}

\usepackage[utf8]{inputenc} 
\usepackage[T1]{fontenc}    
\usepackage{hyperref}       
\usepackage{amsmath}
\usepackage{url}            
\usepackage{booktabs}       
\usepackage{amsfonts}       
\usepackage{nicefrac}       
\usepackage{microtype}      
\usepackage{lipsum}
\usepackage{fancyhdr}       
\usepackage{graphicx}       
\usepackage{epstopdf}
\graphicspath{{media/}}     

\title{Enhanced UV Detection in GaN-Based Photodetectors through InN/AlN  Heterostructure Integration and Doping-Engineered PIN Architecture
\thanks{\textit{\underline{Citation}}: 
\textbf{Authors. Title. Pages.... DOI:000000/11111.}} 
}

\author{Mustafa Kilin$^{a}$$^{*}$, Orkun Tanriverdi$^{a}$, Buldan Karahan$^{b}$, Firat Yasar $^{c}$ \\
        \small $^{a}$Adiyaman University, Adiyaman, Turkiye \\
        \small $^{b}$Hasan Kalyoncu University, Gaziantep, Turkiye \\
        \small $^{c}$Jet Propulsion Laboratory, NASA, Pasadena, CA 91101 \\
        \small $^{*}$Corresponding author: Mustafa Kilin; \tt{mkilin@adiyaman.edu.tr} \\
}

\begin{document}
\maketitle
\begin{abstract}
This study presents a comprehensive simulation-based optimization of gallium nitride (GaN)-based metal–semiconductor–metal (MSM) photodetectors designed for ultraviolet (UV) applications. The proposed device architecture incorporates a novel indium nitride/gallium nitride/aluminum nitride (InN/GaN/AlN) heterostructure integrated on a sapphire substrate, combined with refined doping strategies and interdigitated electrode geometry. By systematically analyzing the effects of mesa layer thickness, buffer layers, substrate type, and doping concentrations, we demonstrate significant enhancements in photocurrent generation, photoabsorption rate, and spectral responsivity. Notably, replacing the conventional sapphire substrate with silicon carbide (SiC) and introducing low-level p-type and n-type (p–n) doping into the GaN region enables p–i–n diode-like behavior, contributing to reduced dark current and improved UV selectivity. Building upon these structural enhancements, the final geometric optimization of the nickel/gold (Ni/Au) electrode fingers led to an approximately 8-fold increase in photocurrent compared to the initial design, representing the most significant contribution to the improvement in absorption efficiency. These findings offer an effective route for designing next-generation MSM photodetectors with improved sensitivity, noise performance, and thermal compatibility, suitable for high-performance ultraviolet detection applications.
\end{abstract}




\section{\label{sec:Introduction}Introduction}

is a widebandgap semiconductor material that has demonstrated remarkable potential for optoelectronic applications, particularly in the UV range \cite{morkocc2002gan, zhang2009low,  zaidi2013highly, yasar2025gan, monroy1998gan}. Its direct band gap of approximately 3.4 eV enables intrinsic UV sensitivity without the need for additional optical filtering \cite{van1997ultraviolet, Strite1992gan}. Additionally, GaN exhibits excellent thermal stability, high carrier mobility, and strong radiation hardness, making it suitable for high-performance UV photodetectors \cite{cai2021progress, Fan2020Analysis, nallabala2021high}. MSM photodetectors are popular because they are easy to fabricate, work at high speeds, and are compatible with large-scale integration \cite{yasar2018flexible, chu2022improved, vivien2008metal}. However, optimizing the design parameters of GaN MSM photodetectors is crucial for maximizing their performance. This includes fine-tuning the thickness of the GaN active layer \cite{alhelfi2016simulation}, optimizing the geometry of the metal fingers \cite{jain2018effect, li2021enhanced}, exploring alternative active layer structures \cite{averine2008solar, chuah2008high, hamdoune2014performance}, doping the GaN thin film \cite{Hamza2017efficient, Steckl2002rare}, and selecting the most suitable substrate for device fabrication \cite{lin1993comparative, chuang2024comparison, chuah2008large}. The thickness of the GaN active layer significantly influences light absorption, carrier transport, and device efficiency. A thicker GaN layer absorbs more light and improves responsivity but slows down the response \cite{Fang2023breaking}.  A very thin GaN layer, on the other hand, does not absorb enough UV light, reducing the photocurrent \cite{Haiping2018simulation}. Thus, an optimized GaN thickness is essential for balancing responsivity and speed. The metal electrode design, including finger width, spacing, and aspect ratio, incorporates these components. A reduced finger spacing enhances the electric field, improving carrier drift velocity and response speed, but may also increase dark current due to tunneling effects \cite{Averina2001Geometry}. Additionally, the choice of Schottky contact metals affects the barrier height, influencing the noise performance of the device \cite{Gu2013Ohmic}.

The optimization of the geometric design of metal fingers is critical to achieving an optimal balance between high-speed operation and low-noise performance \cite{Averine2000Optimization}. To enhance the optoelectronic performance of UV photodetectors beyond geometrical optimization, AlN/GaN/InN multilayer heterostructures have been strategically engineered to exploit their complementary band gaps and polarization-induced electric fields, enabling efficient carrier generation and transport \cite{XINJING2019Structural}. The wide band gap and strong spontaneous polarization of AlN (~6.2 eV) serve to confine carriers and suppress leakage currents, while the intermediate GaN layer (~3.4 eV) functions as the primary absorption region \cite{GEDAM201559}. Finally, an ultrathin InN interlayer (~0.7 eV) introduces strong interfacial electric fields and energy band bending, significantly enhancing carrier separation and responsivity in the deep-UV regime \cite{Paul2014}. Similarly, GaN nanowires and quantum dots increase the active surface area, enhancing light absorption and carrier collection efficiency \cite{Almalawi2020Enhance}. The choice of substrate affects the material quality, strain effects, and overall device performance. Commonly used substrates for GaN growth include sapphire and SiC \cite{GURNETT2006Native, LUO2016Heat}. Sapphire is cost-effective and optically transparent in the UV range, but suffers from a high lattice mismatch with GaN, leading to dislocations \cite{Pharkphoumy2023Correlation}. SiC, on the other hand, provides better lattice matching and superior thermal conductivity, making it a preferred choice for high-power and high-temperature applications \cite{Vittorio2024High}. The optimization of GaN MSM photodetectors involves a multifaceted approach, including material selection, structural design, and substrate engineering. By carefully tuning these parameters, significant improvements in responsivity, speed, and noise characteristics can be achieved. This work presents a systematic optimization strategy aimed at enhancing the performance of GaN MSM photodetectors, contributing to the advancement of high-efficiency UV photodetection technologies.

\section{\label{sec:Material and Design}Material and Design}

The performance of MSM photodetectors strongly depends on the coordinated selection of materials and structural design parameters \cite{Robinson2022Novel}. Table 1 summarizes key physical properties such as lattice parameters, band gap energies, and thermal conductivities of the semiconductors and metals used in this study. These parameters collectively affect the electrical field distribution, optical absorption, and thermal stability of the device.

\begin{table*}
\centering
\caption{The table summarizes key material properties of wide-bandgap semiconductors (e.g., GaN, SiC, AlN) and metals (e.g., Ni, Au) used in MSM detectors. Wide-bandgap semiconductors provide high breakdown voltage and low dark current, while metals enhance carrier transport as Schottky contacts \cite{Strite1992gan},\cite{Kittel2005,Zhang2024}}
\begin{tabular}{|c|c|c|c|}
\hline
\textbf{Material} & \textbf{Lattice Parameters (\AA)} & \textbf{Bandgap (eV)} & \textbf{Thermal Conductivity (W/m·K)}  \\ \hline
Sapphire (Al$_2$O$_3$) & $a = 4.76$, $c = 12.99$ & Transparent in UV & 35  \\ \hline
SiC (4H-SiC) & $a = 3.08$, $c = 15.12$ & 3.2 & 120–180 y \\ \hline
GaN & $a = 3.189$, $c = 5.185$ & 3.4 & 130  \\ \hline
InN & $a = 3.54$, $c = 5.70$ & 0.7 & 45  \\ \hline
AlN & $a = 3.11$, $c = 4.98$ & 6.2 & 285  \\ \hline
Ni & $a = 3.52$ & Metal & 90  \\ \hline
Au & $a = 4.08$ & Metal & 317 \\ \hline
\end{tabular}
\label{tab:materials}
\end{table*}

GaN is employed as the main active layer due to its direct bandgap, which provides strong UV absorption with relatively high electron mobility~\cite{CHEN2020100578}. To improve lattice compatibility and reduce defect-related recombination, AlN is used as a buffer layer, while InN is integrated as an ultra-thin (about \( x < 0.005\,\mu\mathrm{m} \)) interlayer to enhance carrier confinement~\cite{Tsao2018UWBG}. These III-nitride combinations also allow partial band engineering, which contributes to more efficient carrier separation without significantly increasing complexity.

The choice of substrate either sapphire or SiC directly influences both lattice matching and thermal performance~\cite{Vur2001}. While sapphire offers transparency and cost-effectiveness, SiC provides superior thermal conductivity and mechanical robustness, making it favorable for high-power or elevated-temperature applications.

Doping concentration is another critical factor that modulates carrier dynamics in MSM photodetectors. Within the proposed AlN/GaN/InN heterostructure, the AlN layer typically insulating can be engineered with light \textit{p}-type GaN doping to enhance hole transport and serve as an effective buffer for carrier separation. Conversely, the InN region can benefit from \textit{n}-type GaN doping, optimizing electron mobility while mitigating recombination losses. This selective doping strategy not only facilitates efficient charge separation but also refines the internal electric field profile. Stepwise optimization of doping concentrations, ranging from $1 \times 10^{1}$ to $1 \times 10^{19}~\mathrm{cm}^{-3}$,  has been shown to significantly influence both charge transport characteristics and the spatial distribution of the electric field~\cite{Zarate2023optimization}.

Finally, the geometry and material of the metal electrodes play a decisive role in determining device responsivity and speed. Interdigitated finger structures are optimized in width, spacing, and thickness to control electric field intensity and carrier transit times~\cite{jain2018effect,li2021enhanced}. To achieve suitable Schottky barrier heights and stable contact behavior, Ni/Au is selected as the electrode pair due to its well-established work function alignment with GaN~\cite{Guven2004Schotty}.

The schematic representation of the proposed MSM photodetector structure is detailed in Figure 1. As illustrated in Figure 1a, the device operates based on a back-to-back Schottky contact configuration, which leads to charge accumulation at the metal--semiconductor interfaces. The corresponding design of the interdigitated metal fingers, including the dimensions and arrangement of the floating Au and Ni layers, is presented in Figure 1b. Furthermore, Figure 1c compares the final device architectures fabricated on sapphire and SiC substrates, both incorporating a mesa structure formed by AlN, GaN, and an ultrathin InN interlayer. These visualizations collectively support the material and structural choices discussed in this section.

This integrated material and design approach aims not to fully eliminate all limitations, but to address key performance bottlenecks with coordinated structural improvements across the device architecture.

\begin{figure}
\centering
\includegraphics[width=0.48\columnwidth]{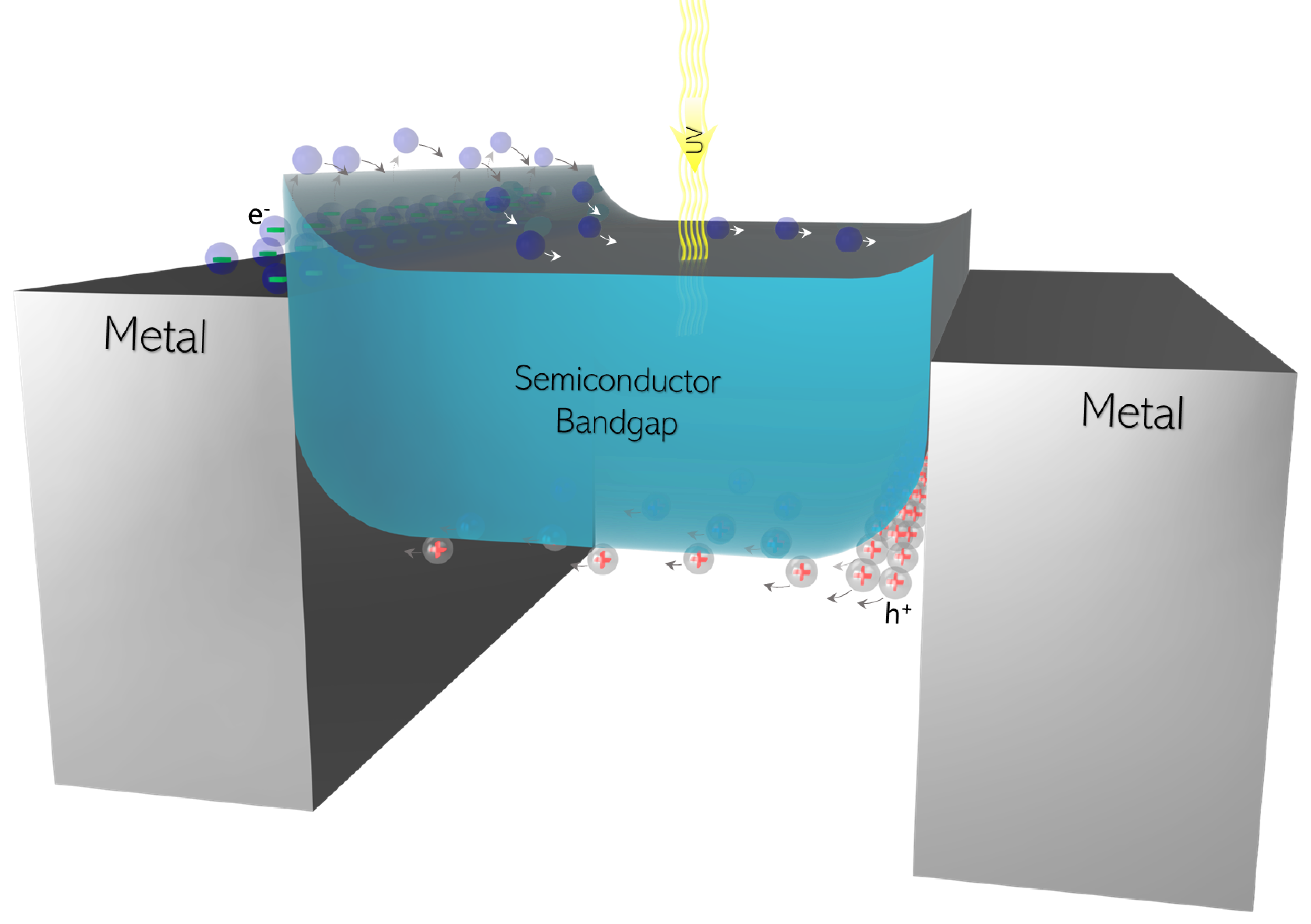}
\hfill
\includegraphics[width=0.48\columnwidth]{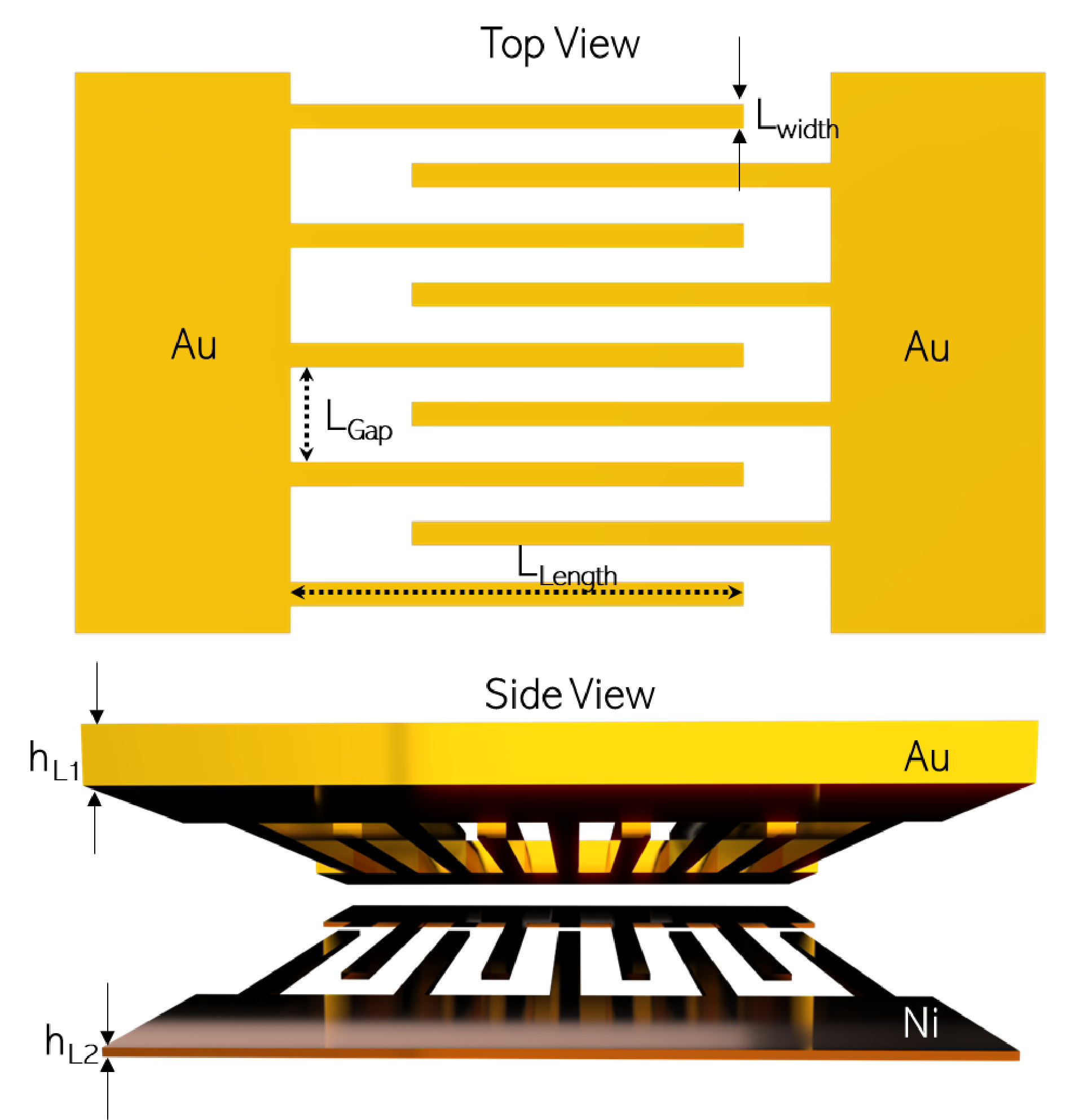}

\vspace{0.5ex}
\parbox[t]{0.48\columnwidth}{\centering (a)}
\hfill
\parbox[t]{0.48\columnwidth}{\centering (b)}

\vspace{2ex}

\includegraphics[width=0.98\columnwidth]{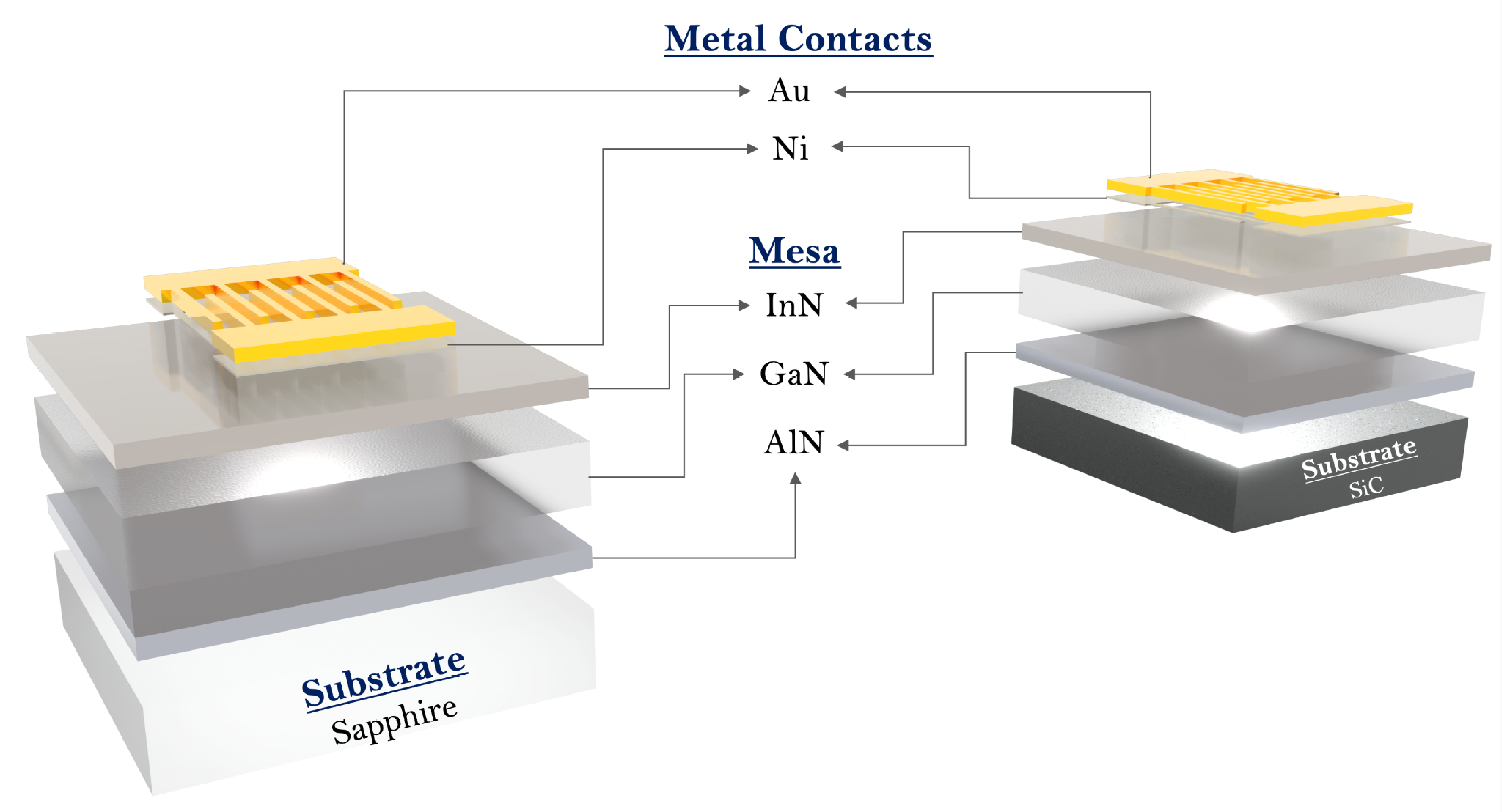}

\vspace{0.5ex}
\parbox[t]{0.98\columnwidth}{\centering (c)}

\caption{\label{fig:fig1}Visual representation of the MSM photodetector: (a) Band diagram showing back-to-back Schottky barriers. (b) Metal finger structure of Au/Ni thin films. (c) Fabricated device layout on sapphire and SiC substrates.}
\end{figure}

\section{\label{sec:Simulations and Methodology}Simulations and Methodology}

The simulation process was conducted using the Silvaco Atlas TCAD environment to investigate and optimize the performance of GaN-based MSM UV photodetectors. All models were constructed in two dimensions under steady-state conditions at room temperature (300 K), solving the Poisson equation and carrier continuity equations self-consistently. The physical models included drift-diffusion transport, thermionic emission at Schottky contacts, and field-dependent mobility effects. Each simulation step focused on enhancing one or more performance metrics such as photocurrent, photoabsorption rate, or spectral responsivity.

Initially, a baseline structure consisting of a sapphire substrate, GaN active region, and Ni/Au electrodes was modeled to evaluate the impact of GaN mesa thickness. The GaN layer thickness was varied from 0.1 $\mu\text{m}$ to 1.0 $\mu\text{m}$, and the optimal absorption rate was obtained at \( h_{GaN} =0.4 \) $\mu\text{m}$, where the trade-off between light penetration depth and carrier transport was best balanced.

Subsequently, an AlN buffer layer  and an almost ultrathin InN surface layer were integrated into the mesa to form a novel AlN/GaN/InN heterostructure. While the AlN thickness was fixed at $h_{\text{AlN}} = 0.1\ \mu\text{m}$, the InN layer thickness was varied from $0.01\ \mu\text{m}$ to $0.1\ \mu\text{m}$, with maximum photoabsorption observed at $h_{\text{InN}} = 0.01\ \mu\text{m}$. This configuration exploited the polarization-induced electric field at the heterointerfaces, leading to improved carrier separation and internal electric field modulation. The AlN layer served as a strain-relieving buffer between sapphire and GaN, improving crystalline quality without completely eliminating mismatch-related defects.

To evaluate the substrate effect, sapphire was replaced by SiC in an identical layer configuration. The SiC-based design demonstrated enhanced thermal conductivity and photocurrent under identical bias conditions, offering better high-temperature stability and increased UV response.

Following the heterostructure optimization, doping profiles were introduced into the GaN region to induce p–i–n diode-like behavior in the MSM configuration. A three-layer GaN stack was created with symmetric p-type and n-type doping around an undoped GaN core. Doping concentrations were swept from  \(1 \times 10^1\) cm\(^{-3}\) to  \(1 \times 10^{19}\) cm\(^{-3}\) , with optimal values of  \( doping_{p-type}=1\times 10^5\) cm\(^{-3}\) for p-type and  \(doping_{n-type}=1 \times 10^{12}\) cm\(^{-3}\) for n-type layers. This configuration enhanced the internal electric field and reduced the dark current while maintaining high responsivity.

Lastly, the interdigitated electrode geometry was refined to maximize the absorption and carrier collection efficiency. The Ni layer thickness was fixed at \( h_{Ni} =0.01\) $\mu\text{m}$ while the Au layer thickness was varied from 0.1 $\mu\text{m}$ to 0.2 $\mu\text{m}$. The highest absorption was recorded at \( h_{Au} =0.12\)  $\mu\text{m}$ Au thickness. Finger widths and gaps were then co-optimized, with the best configuration determined as  \( L_{Width} = 8.0 \) $\mu\text{m}$  and \( L_{Gap} = 19 \) $\mu\text{m}$. All geometrical and material variations were evaluated based on their impact on current-voltage (I–V) characteristics, absorption profiles, and UV spectral responsivity.

Throughout the simulation, fine meshing was applied around the heterojunctions and metal-semiconductor interfaces to ensure numerical stability and accurate field distribution. The results were verified across multiple bias conditions to confirm robustness. While the simulations predict substantial performance improvements, potential discrepancies arising from interface defects or fabrication limitations are acknowledged and considered in the discussion.

\section{\label{sec:Results and Discussion}Results and Discussion}

This section presents a comparative analysis of the simulated performance metrics obtained from a series of sequential structural optimizations applied to GaN-based MSM UV photodetectors. All design variants were modeled using Silvaco Atlas TCAD and evaluated based on their impact on key figures of merit, including photocurrent generation, photoabsorption rate, and spectral responsivity. For clarity and brevity throughout this section, each configuration is denoted with a specific abbreviation: MSM-Ref (baseline sapphire/GaN/Ni/Au structure with optimized mesa thickness), MSM-Buffer (sapphire/AlN/GaN/InN/Ni/Au incorporating AlN and InN buffer layers), MSM-SiC (SiC/AlN/GaN/InN/Ni/Au employing a SiC substrate), MSM-PIN (sapphire/p-GaN/GaN/n-GaN/Ni/Au introducing p–i–n-like doping), and MSM-Opt (final design with optimized Ni/Au electrode geometry). Rather than reporting results in isolation, the analysis focuses on relative comparisons between these configurations under consistent biasing conditions, thereby elucidating both the individual contributions and cumulative enhancements introduced at each design stage. This approach enables a comprehensive understanding of how systematic structural modifications influence the overall photodetector performance.

\subsection{\label{sec:MSM-Ref : sapphire/GaN/Ni/Au}MSM-Ref : sapphire/GaN/Ni/Au}

The initial simulation phase focused on the reference MSM photodetector structure composed of a GaN active layer (\( h_{GaN} =0.4 \) $\mu\text{m}$) deposited on a sapphire substrate with Ni/Au interdigitated electrodes. As illustrated in Figure 2a, the I–V characteristics under UV illumination (360 nm) exhibit a pronounced asymmetry due to the back-to-back Schottky barrier configuration. The photocurrent reaches a peak value of approximately $2.74 \times 10^{-8}$ A at +3 V, while the dark current remains significantly lower, around $5.73 \times 10^{-14}$ A, outperforming the results reported in \cite{Zhang2025Novel}, resulting in a photo-to-dark current ratio (PDCR) exceeding $1.74 \times 10^7$ (Figure 2b), exhibiting a higher PDCR than that reported in  \cite{Butun2008High, Wen2024High}. This high ratio indicates strong signal-to-noise characteristics and effective suppression of leakage pathways. The spectral responsivity, shown in Figure 2c, remains centered around 360 nm with an amplitude consistent with the intrinsic absorption edge of GaN. At this wavelength, a responsivity of 0.5275 A/W is achieved, validating the baseline structure and establishing a quantitative foundation for evaluating subsequent heterostructural and geometrical optimizations. This responsivity value notably surpasses that reported by Butun et al. , who achieved 0.23 A/W at 356 nm under similar bias conditions \cite{Butun2008High}. Additionally, the photon absorption rate (par) in the GaN region reaches approximately $1.51 \times 10^{24}$ $ par / cm^{3} . s $, as determined by the optical intensity distribution across the mesa geometry. This value reflects the strong confinement and penetration of incident UV radiation within the active layer, reinforcing the effectiveness of the vertical photogeneration profile in this baseline configuration.

Building upon the baseline performance, MSM-Ref configuration establishes a reliable reference point for assessing further enhancements. While the results demonstrate excellent dark current suppression and UV responsivity, certain intrinsic limitations remain—particularly in terms of strain management, electric field modulation, and carrier separation efficiency. To address these aspects, the following device iterations explore the inclusion of an AlN buffer to alleviate lattice mismatch, the use of an ultrathin InN interlayer to enhance internal electric fields, the substitution of sapphire with SiC for improved thermal and structural compatibility, and the introduction of p-type and n-type doping to induce p–i–n diode-like characteristics. Electrode geometry is also refined to optimize optical absorption and charge collection. These progressive modifications are examined in detail in the subsequent sections, each designed to build upon the foundation established by the reference device.

\begin{figure}
\centering
\includegraphics[width=0.48\columnwidth]{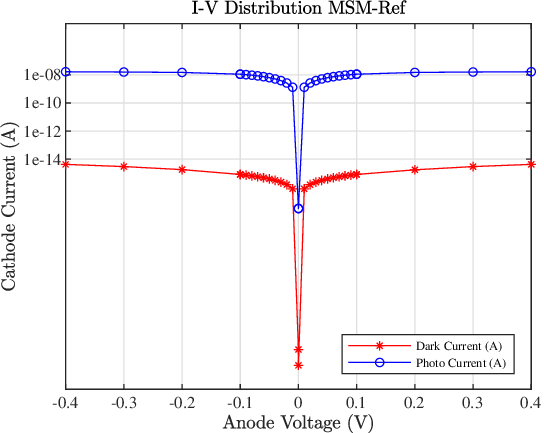}
\hfill
\includegraphics[width=0.48\columnwidth]{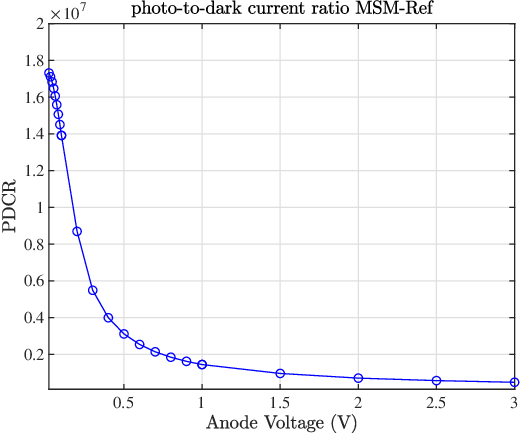}

\vspace{0.5ex}
\parbox[t]{0.48\columnwidth}{\centering (a)}
\hfill
\parbox[t]{0.48\columnwidth}{\centering (b)}

\vspace{2ex}

\includegraphics[width=0.98\columnwidth]{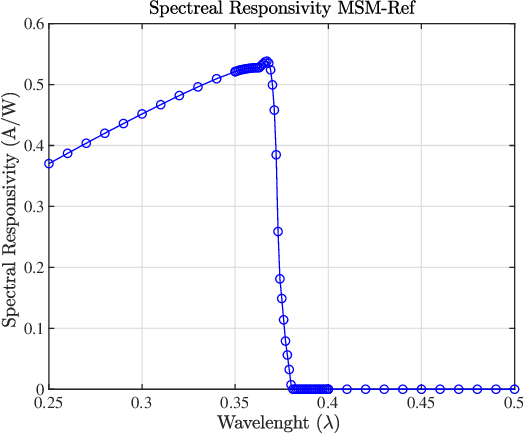}

\vspace{0.5ex}
\parbox[t]{0.98\columnwidth}{\centering (c)}

\caption{\label{fig:fig2}In the MSM-Ref detector structure for the optimized GaN layer thickness of 0.4 $\mu\text{m}$ , (a) I-V distrubition,  (b) PDCR,  (c) spectral responsivity A/W.}
\end{figure}

\subsection{MSM-Buffer:Sapphire/AlN/GaN/InN/Ni/Au}
\subsection{MSM-SiC : SiC/AlN/GaN/InN/Ni/Au}

In order to extend the MSM-Ref design, structural variations were introduced by incorporating a mesa region composed of an AlN/GaN/InN heterostructure, followed by replacing the sapphire substrate with SiC to assess substrate-induced effects. The design-related optimization processes were detailed in the Simulations and Methodology section.  The first configuration MSM-Buffer yielded a peak photocurrent of $3.23 \times 10^{-7}$ A and a dark current of $2.69 \times 10^{-14}$ A, corresponding to a PDCR of $1.97 \times 10^{7}$, a photoabsorption rate of approximately $1.5 \times 10^{24}$ $ par / cm^{3} . s $, and a responsivity of 0.6224 A/W. Compared to the MSM-Ref, these results reflect an enhancement in responsivity and a further suppression of dark current, attributable to the InN interlayer's modulation of the internal electric field and the AlN buffer's role in lattice strain mitigation. In contrast, the second configuration, MSM-SiC, while benefitting from improved lattice matching and thermal conductivity, showed a lower peak photocurrent of $3.09 \times 10^{-8}$ A and a higher dark current of $2.68 \times 10^{-13}$ A. Consequently, the PDCR dropped to $3.5 \times 10^{5}$, and responsivity decreased to 0.4463 A/W, with the photoabsorption rate settling around $7 \times 10^{22}$ $ par / cm^{3} .s$ These performance shifts suggest that while SiC substrates enhance thermal robustness, the concurrent modifications to the electric field distribution and carrier dynamics may limit photocurrent gain in this heterostructure context. Nevertheless, both configurations demonstrate structurally tunable performance, forming a foundation upon which additional enhancements such as doping-induced p-i-n diode-like profiles and electrode geometry optimization can be systematically built.

\subsection{MSM-PIN : sapphire/$p_{GaN}$/GaN/$n_{GaN}$/Ni/Au}

Since the anticipated photocurrent enhancement in the MSM-SiC structure engineered by replacing the sapphire substrate with SiC was found to be limited primarily to thermal effects, subsequent structural development focused on improving the MSM-Buffer configuration. To this end, a p–i–n diode-like architecture was introduced through strategic doping profile engineering, aiming to establish a built-in electric field and thereby enhance carrier separation and overall device responsivity. A symmetric doping profile was implemented on the MSM-Buffer heterostructure by introducing p-type and n-type GaN layers surrounding an intrinsic GaN core, as described in the section 3. This modified device configuration, MSM-PIN : sapphire / p-GaN / GaN / n-GaN / Ni / Au, produced a photocurrent of $3.88 \times 10^{-7}$ A and a dark current of $3.24 \times 10^{-14}$ A, resulting in a PDCR of $2.35 \times 10^{7}$ and a photoabsorption rate of approximately $2.2 \times 10^{24}$ $ par / cm^{3} . s $. A responsivity of 0.6605 A/W was achieved, surpassing all previously examined structures in both magnitude and spectral sharpness. These improvements can be attributed to the built-in electric field across the p-i-n profile, which enhances carrier separation efficiency and reduces recombination losses. As illustrated in Figure~3a, the PDCR of the doped structure closely approaches that of the best-performing undoped design, while offering improved stability under bias. Additionally, the spectral responsivity curve shown in Figure~3b confirms a sharper UV selectivity with minimal response leakage into longer wavelengths. Considering all four configurations examined, namely MSM-Ref, MSM-Buffer, MSM-SiC, and MSM-PIN demonstrates the most balanced trade-off between signal strength, noise suppression, and spectral precision. These findings highlight the effectiveness of doping-based internal field modulation as a reliable and fabrication-friendly approach for advancing GaN MSM photodetector performance.

\begin{figure}[!t]
\centering

\includegraphics[width=0.48\columnwidth]{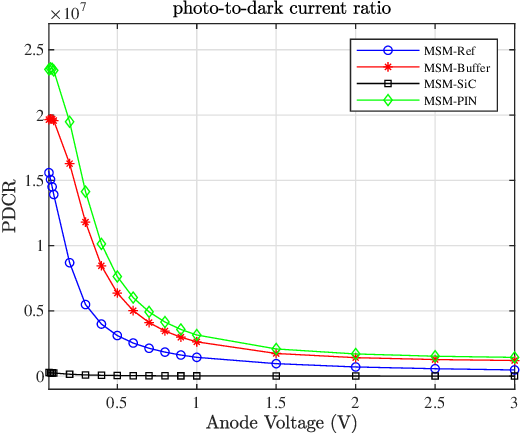}
\hfill
\includegraphics[width=0.48\columnwidth]{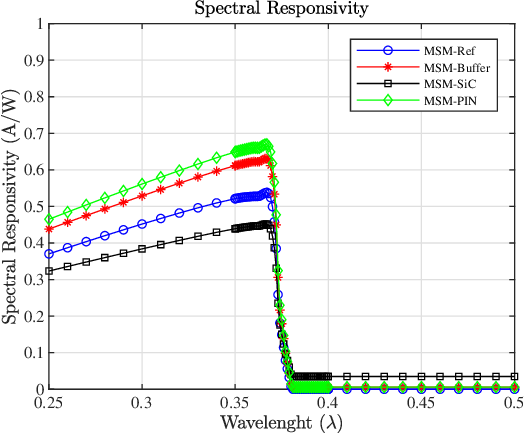}

\vspace{0.5ex}
\parbox[t]{0.48\columnwidth}{\centering (a)}
\hfill
\parbox[t]{0.48\columnwidth}{\centering (b)}

\caption{Presents a comparative analysis of four MSM design a) PDCR b) spectral responsivity}
\label{fig3}
\end{figure}

As illustrated in Figure~3a, PDCR values across the investigated structures demonstrate substantial improvement compared to the conventional MSM photodetectors reported in the literature. In particular, the doped MSM-PIN configuration exhibits a PDCR that exceeds $2 \times 10^{7}$, making it one of the most performing designs when compared with recent studies such as Butun et al.~\cite{Butun2008High} and Wen et al.~\cite{Wen2024High}, both of which report PDCR values below $10^{6}$ under similar bias conditions. Likewise, the spectral responsivity distribution shown in Figure~3b confirms the effectiveness of the proposed structural strategies to improve the sensitivity to UV detection while maintaining narrowband selectivity. A peak responsivity of 0.6605 A/W at 360 nm surpasses previously published values for GaN-based MSM devices without internal amplification. These results collectively underscore the success of the employed heterostructural and doping-based modifications in achieving high signal-to-noise performance and optical selectivity. Building upon these improvements, the final stage of optimization focuses on refining the geometry of the interdigitated electrode, with the goal of further improving absorption efficiency and carrier collection across the active region.

\subsection{MSM-Opt : sapphire/$p_{GaN}$/GaN/$n_{GaN}$/Ni/Au with electrot optimization}

Following the structural and doping-based enhancements, the final optimization stage focused on refining the electrode geometry to improve optical absorption and charge collection efficiency. As detailed in section 3, variations in metal thickness, finger width, and spacing were evaluated to maximize the electric field uniformity across the active GaN region. The optimized design based on the MSM-PIN configuration achieved a peak photocurrent of $2.36 \times 10^{-7}$ A and a dark current of $1.84 \times 10^{-14}$ A, resulting in an exceptional PDCR of $2.52 \times 10^{8}$ and a responsivity of 0.8159 A/W. The photoabsorption rate reached $2.25 \times 10^{24}$ $ par / cm^{3} . s $, indicating effective confinement of incident radiation within the mesa region. These results represent the highest responsivity and signal-to-noise performance recorded across all evaluated structures, surpassing even the doped p-i-n diode-like device described previously. Such improvements can be directly attributed to the optimized Ni/Au finger dimensions, which promote enhanced optical penetration and lateral carrier drift while minimizing electrode-induced shadowing. These findings underscore the critical role of geometrical design in complementing material and doping strategies for GaN-based MSM photodetectors. A detailed comparison of this geometry optimized device with the best performing doped structure is presented in Figure~4.

\begin{figure}[!t]
\centering

\includegraphics[width=0.48\columnwidth]{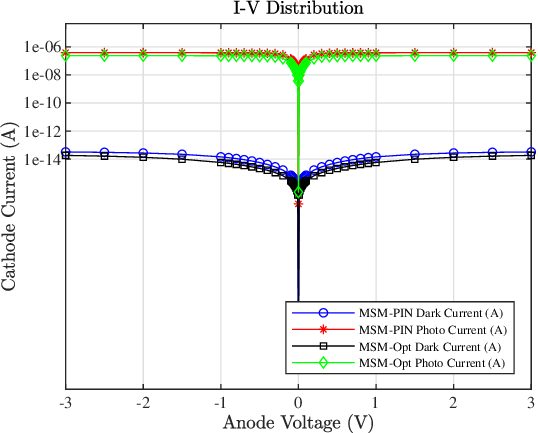}
\hfill
\includegraphics[width=0.48\columnwidth]{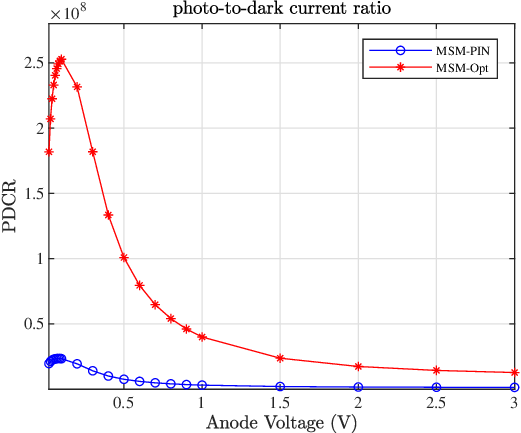}

\vspace{0.5ex}
\parbox[t]{0.48\columnwidth}{\centering (a)}
\hfill
\parbox[t]{0.48\columnwidth}{\centering (b)}

\vspace{2ex}

\includegraphics[width=0.98\columnwidth]{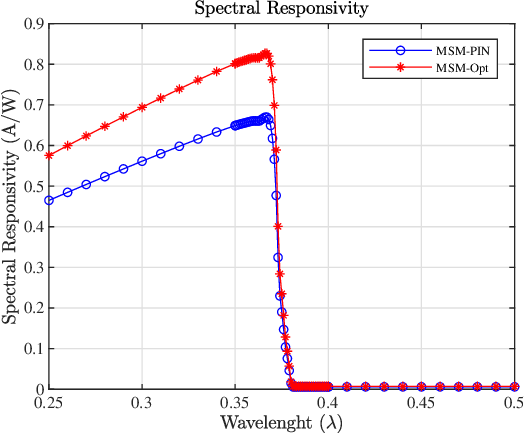}

\vspace{0.5ex}
\parbox[t]{0.98\columnwidth}{\centering (c)}

\caption{Comparison of the two best-performing MSM photodetector configurations: the sapphire/p-GaN/GaN/n-GaN/Ni/Au structure and the geometry-optimized sapphire/AlN/GaN/InN/Ni/Au structure. 
(a) Current–voltage (I–V) characteristics under UV illumination (360 nm), highlighting enhanced carrier transport in the geometry-optimized device at higher bias.
(b) PDCR across both structures, showing superior signal-to-noise performance in the geometry-optimized configuration.
(c) Spectral responsivity comparison, indicating improved responsivity and sharper UV selectivity for the electrode-optimized design.}
\label{fig4}
\end{figure}

Figure 4 provides a comprehensive comparison between the two most advanced device architectures: the MSM-PIN structure and its geometrically optimized counterpart, MSM-Opt, which incorporates refined electrode design for enhanced performance. As shown in Figure 4a, the geometry-optimized device exhibits a slightly lower photocurrent than the doped configuration at low bias, but outperforms it under higher bias conditions, indicating superior field-assisted carrier transport. In Figure 4b, the PDCR of the geometry-optimized structure surpasses that of the doped device by more than an order of magnitude, highlighting the advantage of minimal leakage current combined with efficient photoresponse. Figure 4c further confirms this trend, showing a peak responsivity of 0.8159 A/W for the optimized geometry, compared to 0.6605 A/W for the doped device. These comparisons reveal that while doping strategies significantly enhance internal electric field modulation, precise control of electrode architecture can yield even greater performance gains. Therefore, electrode geometry should be considered not as a secondary adjustment but as a principal design axis in high-performance MSM photodetector development.

\section{\label{sec:Conclusion}Conclusion}

In this study, a comprehensive simulation-based optimization of GaN-based MSM photodetectors was conducted using Silvaco Atlas TCAD to enhance UV detection performance. Starting from a baseline MSM-Ref configuration, several structural and material modifications were systematically implemented and analyzed. The introduction of an MSM-Buffer heterostructure significantly improved photoabsorption and carrier separation, while the use of a SiC substrate demonstrated thermal advantages but at the cost of reduced responsivity. Further enhancements were achieved through the integration of symmetric p-type and n-type doping around the intrinsic GaN region, forming a p-i-n diode-like structure that increased internal electric field strength and reduced recombination losses. The final optimization, involving precise adjustments to Ni/Au electrode geometry, resulted in the best overall performance, yielding a PDCR of $2.52 \times 10^{8}$ and a peak responsivity of 0.8159 A/W.

\begin{figure}[!t]
\centering

\includegraphics[width=0.48\columnwidth]{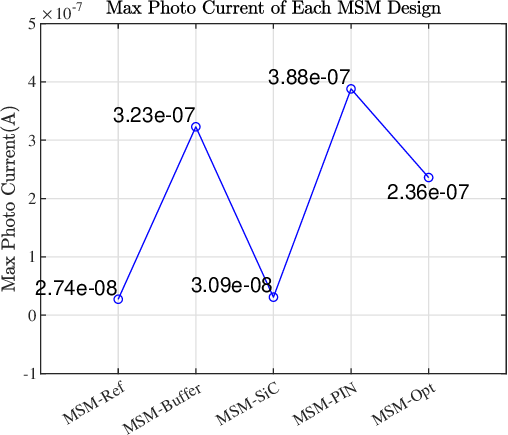}
\hfill
\includegraphics[width=0.48\columnwidth]{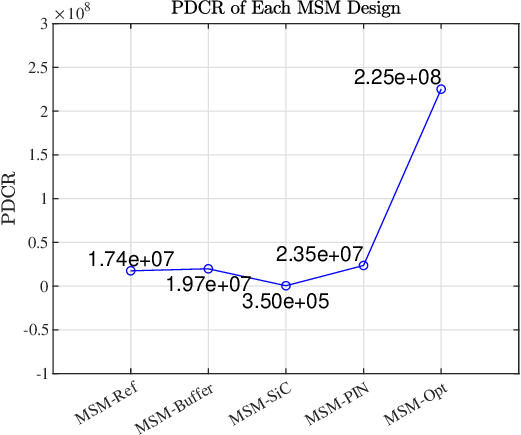}
\vspace{0.5ex}
\parbox[t]{0.48\columnwidth}{\centering (a)}
\hfill
\parbox[t]{0.48\columnwidth}{\centering (b)}

\vspace{2ex}

\includegraphics[width=0.48\columnwidth]{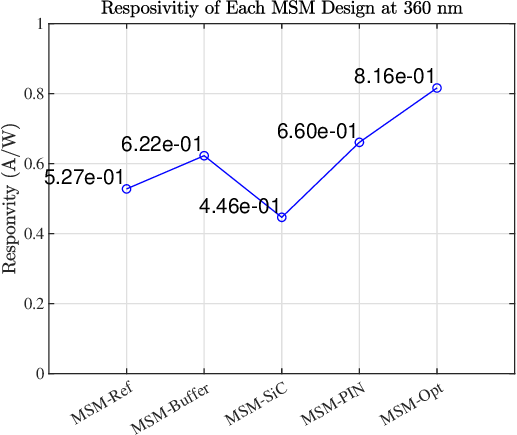}
\hfill
\includegraphics[width=0.48\columnwidth]{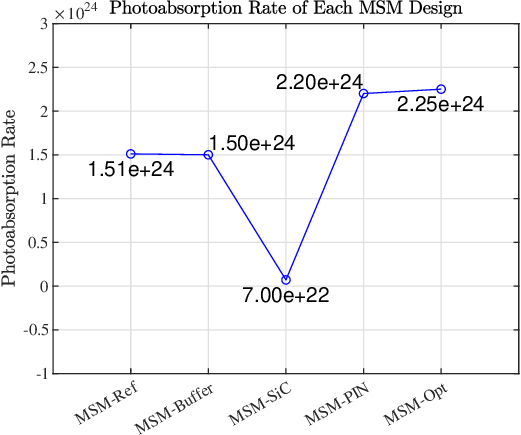}
\vspace{0.5ex}
\parbox[t]{0.98\columnwidth}{\centering (c)}
\hfill
\parbox[t]{0.48\columnwidth}{\centering (d)}
\caption{Normalized performance metrics of five distinct MSM photodetector structures are presented to highlight the impact of sequential design optimizations. The graphs show (a) the maximum photocurrent under +3 V bias, (b) PDCR, (c) the spectral responsivity at 360 nm, and (d) the photoabsorption rate at +3 V. These comparative results reflect the effectiveness of structural improvements including buffer layer engineering, substrate replacement with SiC, doping-induced internal electric field enhancement via p–i–n-like configurations, and electrode geometry refinement, each contributing incrementally to enhanced UV detection performance.}
\label{fig5}
\end{figure}

Figure 5 presents a consolidated overview of the key performance metrics corresponding to each MSM photodetector configuration. Figure 5a displays the photocurrent characteristics under forward bias conditions ranging from 0 to +3 V, demonstrating a steady improvement in current levels through successive design enhancements. Figure 5b illustrates the maximum photo-to-dark current ratio (PDCR) measured at +3 V, where the MSM-PIN and MSM-opt structures exhibit superior signal-to-noise performance. In Figure 5c, the spectral responsivity at 360 nm is compared across all configurations, with the highest value achieved in the final geometrically optimized device. Figure 5d shows the corresponding photoabsorption rate under +3 V, reflecting the effectiveness of buffer layer integration and doping in enhancing photon carrier conversion. Collectively, these results highlight the critical influence of material selection, doping profile, and electrode design in advancing the efficiency and selectivity of GaN-based MSM UV photodetectors.

Comparative analyses across different architectures demonstrated that both doping and electrode design play essential roles in determining the balance between signal amplification, noise suppression, and spectral selectivity. The findings underscore that combining material-level engineering with geometrical refinement offers a powerful pathway toward the development of high-efficiency, low-noise UV photodetectors. These results provide valuable design guidelines for future fabrication efforts and suggest promising directions for the integration of GaN MSM photodetectors into high-performance optoelectronic systems.

Starting from the baseline MSM-Ref structure, each subsequent design iteration was strategically introduced to address specific performance limitations: the MSM-Buffer structure incorporated a transport-enhancing buffer layer; MSM-Doped employed controlled doping profiles to modulate carrier concentration and mobility; MSM-PIN reinforced the internal electric field to promote efficient carrier separation; and finally, MSM-Optimized refined electrode geometry and material distribution for maximal extraction efficiency. Throughout this progression, nonlinear yet consistently upward trends were observed in key performance metrics such as photocurrent, PDCR, and responsivity. In particular, the enhancement in responsivity demonstrates a substantial improvement in the device’s light-to-current conversion efficiency. However, this increase was not proportionally mirrored in the photoabsorption rate, indicating that the number of absorbed photons remained relatively constant while improvements in carrier separation and collection mechanisms became the dominant contributors to performance gains. These findings underscore the fact that optimizing only the optical absorption is insufficient; rather, meticulous control over carrier dynamics, via doping engineering, electric field modulation, and contact design, plays a pivotal role in achieving high-performance photodetectors. Collectively, the results confirm the necessity of a holistic design strategy and demonstrate the transformative impact of multidimensional structural optimization on the overall efficiency of GaN-based MSM photodetectors.

\section*{Acknowledgments}
The authors would like to express their gratitude to Adıyaman University for providing the necessary resources and facilities to conduct this research. Special thanks to my colleagues for their valuable insight and technical support throughout the study. In addition, we appreciate the constructive feedback from the anonymous reviewers, which helped to improve the quality of this manuscript.


\end{document}